\documentclass[12pt]{article}

\usepackage{epsfig}
\usepackage{amsmath}
\usepackage{amssymb}
\usepackage{axodraw}

\providecommand{\beqa}{\begin{eqnarray}}
\providecommand{\eeqa}{\end{eqnarray}}

\def\cG{{\mathcal{G}}}
\def\cL{{\mathcal{L}}}

\def\cO{{\mathcal{O}}}

\def\cV{{\mathcal{V}}}

\begin{document}

\begin{titlepage}
\begin{flushright}
UMD-PP-04-034\\
hep-th/0404001\\
\today
\end{flushright}

\vspace{1cm}
\begin{center}

\baselineskip25pt {\Large\bf On De Sitter Vacua in Strongly Coupled \\
Heterotic String Theory\footnote{\uppercase{T}alk at {\it
\uppercase{SUSY} 2003: \uppercase{S}upersymmetry in the
\uppercase{D}esert}\/, held at the \uppercase{U}niversity of
\uppercase{A}rizona, \uppercase{T}ucson, \uppercase{AZ},
\uppercase{J}une 5-10, 2003. \uppercase{T}o appear in the
\uppercase{P}roceedings.}}

\end{center}
\vspace{1cm}
\begin{center}
\baselineskip12pt {Axel Krause\footnote{E-mail: {\tt
krause@physics.umd.edu}}} \vspace{1cm}

{\it Center for String and Particle Theory,}\\[1.8mm]
{\it Department of Physics, University of Maryland}\\[1.8mm]
{\it College Park, MD 20742, USA}

\vspace{0.3cm}
\end{center}
\vspace*{\fill}

\begin{abstract}
We describe how 4d de Sitter vacua might emerge from 11d heterotic
M-theory. Non-perturbative effects and $G$-fluxes play a crucial
role leading to vacua with F-term supersymmetry breaking and a
positive energy density. Charged scalar matter fields are no
longer massless in these vacua thus solving one of the problems of
the heterotic string. Moreover, interesting dark matter candidates
appear in a natural way.
\end{abstract}

\vspace*{\fill}

\end{titlepage}

\section{Introduction}

There is great interest in connecting M-theory to real cosmology.
On the one hand side there are no high energy experiments testing
M-theory in its generic regime (meaning where all extra dimensions
are much smaller than of inverse TeV-scale size) thus rendering
cosmology an important `experimental' alternative. On the other
hand one should expect that M-theory in its final formulation will
be able to give insight into the origin of inflation and
ultimately should tell us how to cure the big bang singularity.
While the latter problem seems to require first the full
microscopic formulation of M-theory for its answer, it seems that
the former can already be tackled with current local field theory
approximations of M-theory in the form of 11d supergravity.

As a first important step in this direction we have to find robust
mechanisms which allow us to obtain 4d de Sitter spacetimes from
M-theory. As de Sitter spacetimes with large unsuppressed
cosmological constants are relevant during inflation, even the
derivation of de Sitter spacetimes without an accompanied solution
to the cosmological constant problem will be important (the
solution to the cosmological constant problem seems once more to
require a much better understanding of the microscopic M-theory
and its degrees of freedom which might very well be discrete and
finite \cite{DDOF}). In this talk I will focus on heterotic
M-theory \cite{HetM} for two reasons. First, heterotic M-theory
includes M-theory in its bulk and is therefore the more general
starting point. Second, due to the $E_8$ gauge groups on its 10d
boundaries and various phenomenological virtues (see \cite{Nilles}
for a review) this theory seems to be ideal to address cosmology
with realistic matter and gauge fields.

Let's consider therefore heterotic M-theory compactified on
$CY\times S^1/Z_2$ ($CY$=Calabi-Yau threefold) from 11d down to
4d. In order to obtain de Sitter vacua we will have to break
supersymmetry. Preferentially, this should happen spontaneously
through F-term breaking. To this end the inclusion of
non-perturbative effects into the dimensionally reduced effective
4d theory will be important. These effects arise either from open
membrane instantons (OMI) stretching through the bulk between the
two boundaries or from gaugino condensation (GC) on the hidden
boundary \cite{GC} (in more complicated vacua also M5-instantons
wrapping the complete internal CY threefold could be included and
might even be required in order to satisfy the anomaly cancelation
condition \cite{DLOW}). As these effects lead to boundary-boundary
forces (for earlier 11d studies of these see \cite{K1}) they are
natural candidates for a stabilization of the dilaton which in
heterotic M-theory corresponds to the orbifold length $\cL$. Note
that further non-perturbative effects which would be allowed by
M-theory, e.g.~OMIs wrapping supersymmetric 3-cycles on the
internal CY threefold are not compatible with the supersymmetry
preserved by the two boundaries. Therefore since we want to start
with a supersymmetric configuration in 11d the first two
non-perturbative effects are exhausting and indeed have to be
included as they cannot be avoided (as to GC note that in
heterotic M-theory the strong coupling of the hidden gauge group
is not an option in contrast to the weakly coupled heterotic
string).

Stabilizing the orbifold length by means of OMI's has been
considered in \cite{CK2} for the linearized warped background of
\cite{WWarp}. This background solves the 11d gravitino
Killing-spinor equation to linear order in a series expansion in
the warp-factor and it turns out that in the regime where this
approximative background is valid, OMI's are the most dominant
non-perturbative effects \cite{MPS} while GC is exponentially
suppressed against them. It is an important feature of heterotic
M-theory that in general it is inconsistent to set all $G$-fluxes
to zero. For instance the standard embedding of the spin- into the
gauge connection no longer leads to a trivial Bianchi identity.
Consequently for an 11d background which preserves 4d, $N=1$
supersymmetry the $G_{2,2}$ (all indices tangent to the CY) flux
component deforms the background such that the CY volume decreases
along the orbifold from visible towards hidden boundary (one could
also have an increase which however doesn't seem to be
phenomenologically relevant). It turns out that at the level of
the effective 4d potential one can stabilize the orbifold modulus
$\cL$ by balancing OMIs against this non-trivial variation of the
background geometry along the orbifold which is generated by the
$G_{2,2}$ flux component \cite{CK2}. However, since the linearized
background exhibits a linearly decreasing warp-factor and also CY
volume one has to introduce an additional M5 brane (4d
spacetime-filling and wrapping an internal holomorphic 2-cycle to
preserve supersymmetry) whose additional $G_{2,2}$ flux
contribution can be used to prevent the metric and therefore the
CY volume from becoming negative. This M5 brane gets then
stabilized in the middle of the interval \cite{MPS}.

\begin{figure}[t]
  \begin{center}
  \epsfig{file=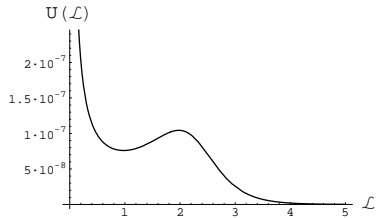,width=4.9cm,angle=0}
  \caption{\it The 4d potential caused by OMI's in units of the
  reduced Planck scale
  as a function of the orbifold length $\cL=L/l$ where $l=2\pi^{1/3}l_{11}$
  and $l_{11}$ is the 11d Planck length.}
  \label{fig1}
  \end{center}
\end{figure}

Unfortunately it turns out that if one wants to study the
effective 4d potential for the orbifold length modulus $\cL$ at
values larger than the stabilized critical one, one enters the
regime where the metric of the linearized background becomes
negative and is therefore no longer Riemannian. To cure this state
of affairs one should go to the exact non-linear background which
always gives a manifestly positive Riemannian metric and therefore
positive CY volume \cite{CK1,CK3}. The linearized background is
recovered as the tangent approximation to the exact solution at
the location of the visible boundary. One then finds working in
the exact background and keeping the parallel M5 brane that by
considering OMI's between the M5 and both boundaries the M5 still
gets stabilized at the middle of the orbifold interval. Moreover
$\cL$ can be stabilized again by a balance between a nontrivial
dependence of the geometry on $\cL$ (due to $G_{2,2}$) and OMI
effects (see fig.\ref{fig1}).
However, huge CY intersection numbers $d\gtrsim 10^4$ (for the
simplest case of a CY with $h^{1,1}=1$) are required. Moreover,
the volume of the OMIs in Planck units, $\cV_{OM}$, turns out to
be smaller than $1$ at the location of the critical $\cL$, thereby
unfortunately showing that at the minimum one looses control over
the supergravity, not to mention that multiply wrapped instantons
are no longer suppressed and would contribute as well.

\begin{figure}[bht]
  \begin{center}
  \epsfig{file=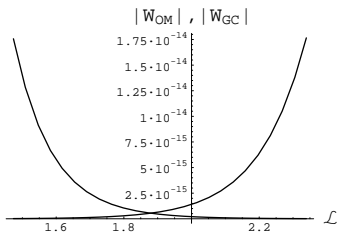,width=4.9cm,angle=0}
  \caption{\it The dependence of the absolute values of the open
               membrane and gaugino condensation superpotentials,
               $|W_{OM}|$ (left curve) and $|W_{GC}|$ (right
               curve) on $\cL$.}
  \label{fig2}
  \end{center}
\end{figure}

The attractive features of the $\cL$ stabilization so far -- {\em
a positive vacuum energy} together with {\em spontaneously broken
supersymmetry due to F-terms} -- can however be kept when one
works in the exact background and takes into account GC
\cite{BCK}. Let us focus here on the simplest case without
additional M5-branes. It is important that in the exact background
there is no longer the need to suppress GC against OMI for
consistency reasoning of the background. The potential due to
OMI's decreases with $\cL$ while that caused by GC increases which
suggests a natural $\cL$ stabilization mechanism by balancing
these two effects against each other (see fig.\ref{fig2}). Indeed
by working out the full 4d effective
potential it turns out
\begin{figure}[th]
  \begin{center}
  \epsfig{file=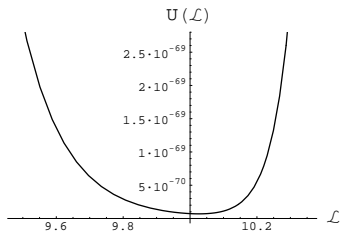,width=4.9cm,angle=0}
  \caption{\it The 4d potential which results from OMI and GC in units
  of the reduced Planck scale as a function of the orbifold length $\cL$.
  The hidden gauge group chosen is $SU(4)$.}
  \label{fig3}
  \end{center}
\end{figure}
that this gives a very robust mechanism of stabilizing $\cL$ which
is equivalent to stabilizing the dilaton (see fig.\ref{fig3}).
Without additional M5-branes the exact background exhibits a naked
singularity at a finite $\cL_{max} = 1/\cG_v$ \cite{CK3} where
$\cG_v$ measures the visible boundary charge associated to the
visible boundary $G_{2,2}$ flux. However, since
$\cG_v\propto\cV_v^{-1/3}$ ($\cV_v$ being the visible boundary
CY-volume) this upper bound on $\cL$ is pushed towards infinity in
the decompactification limit where $\cV_v\rightarrow\infty$. It is
therefore possible to study this limit and to establish the
expected runaway behavior towards a zero energy decompactified
flat space. Consequently our local positive energy de Sitter vacua
are metastable vacua which will not possess the full maximally de
Sitter symmetry. It is therefore likely that their isometry group
has finite-dimensional representations, a point recently stressed
in \cite{MI}.

It is satisfying that the OMI-GC balancing mechanism does not need
very high CY-intersection numbers $d$ anymore but works already
for $d=1$. The stabilized critical $\cL_0$ scales with $d$ as
$\cL_0\propto d^{1/3}$. Moreover, it is essential that now the
critical $\cL_0$ leads to values of the CY volume and OMI volume
which are much bigger than Planck size and therefore show that the
vacua lie in a regime where supergravity is under control, meaning
that higher order corrections to it are sufficiently suppressed.
It turns out that beyond exhibiting a stabilized $\cL$, also the
$S$ and $T$ axions become fixed and the vacuum expectation value
(vev) of the 4d charged matter $C$ fields becomes non-trivial. The
vev attained by the $C$'s acquires an exponential suppression
factor because the minimization for $C$ requires a balancing
between the $C$ vev and the two non-perturbative OMI and GC
effects. It is therefore generic that the $C$ vev lies far below
the (reduced) Planck scale and can be brought close to the TeV
regime. This is interesting as it is one of the problematic
features of 4d heterotic string vacua, next to the runaway of the
dilaton, to give massless charged scalars after supersymmetry
breaking through GC (see however \cite{FP} for a generation of
such mass terms through higher dimensional operators; combined
with an anomalous $U(1)$ these lead to supersymmetry breaking). It
would clearly be interesting to see whether the $C$ vev's could
help breaking the visible GUT groups further down to the standard
model. Note that in heterotic M-theory there is no reason to
prefer the standard embedding of the spin- into the
gauge-connection over any non-standard embedding. One is therefore
not restricted to an $E_6$ GUT group but could also aim to obtain
$SO(10)$ \cite{SO} (which was done in the context of elliptically
fibered CY's e.g.~in \cite{DLOW}) or the Pati-Salam group
$SU(4)_c\times SU(2)_L\times SU(2)_R$ \cite{PS} the two
phenomenologically most favored GUT groups \cite{PR}.

Further moduli like the complex structure moduli and the CY-volume
modulus are expected to get stabilized once a Neveu-Schwarz
$G_{0,3,1}=H_{0,3}$ flux component is switched on and the
respective flux superpotential \cite{BC} $W=\int_{CY}
H_{0,3}\wedge \Omega$ is added. This type of flux, together with
GC and one-loop corrections to the gauge kinetic functions has
e.g.~been used in \cite{GKLM} in the weakly coupled case for the
stabilization of complex structure and K\"ahler moduli. Note that
these `one-loop' corrections appear in heterotic M-theory
automatically at `tree level' and are therefore no longer small.
Moreover, since the $G_{0,3,1}$ flux component is localized by a
delta-function on the boundaries and cannot penetrate the bulk,
the situation is indeed very similar to the weakly coupled case.
It would of course also be interesting to switch on the
$G_{1,2,1}=H_{1,2}$ component leading to non-K\"ahler manifolds
where one would expect a stabilization of the CY-volume modulus at
tree level \cite{NK}. However, in this case one still has to
better understand the moduli structure of these non-K\"ahler
manifolds before one is able to stabilize them.

An interesting property of the resulting de Sitter vacua is the
fact that by choosing the hidden gauge group to be of low rank,
say $SU(4)$ or $SU(3)$ as opposed to an unbroken $E_8$, one can
rather easily bring the supersymmetry breaking scale and gravitino
mass close to the relevant TeV scale \cite{BCK}. In doing so one
stabilizes the hidden boundary close to the maximally allowed
value $\cL_{max}=1/\cG_v$ which is phenomenologically favored as
it leads to the right value for the 4d Newton's Constant once the
Grand Unified gauge coupling and energy scale assume their
standard values \cite{WWarp}, \cite{CK3}. Moreover, the hidden
matter which arises when we have broken the hidden $E_8$ gauge
group down to $SU(4)$ or $SU(3)$ (we take for simplicity simple
groups though product groups with low dual Coxeter number would
qualify as well) say, represents a natural candidate for {\em dark
matter} as it couples to the visible matter only
(super)\-gravitationally and can be expected to enjoy similar
clustering properties required for dark matter to distinguish it
from dark energy. Though the complete vacuum energy turns out to
be exponentially suppressed (similar as in warp-geometries
\cite{KCC}) through the non-perturbative geometrical factors, this
suppression is unfortunately not big enough to bring it down to a
realistic meV vacuum energy scale. What one finds instead confirms
the general expectation, namely that the vacuum energy turns out
to be of the same order as the supersymmetry breaking scale,
though smaller by a factor of $\cO(10)$. Therefore, in the
supergravity approach to de Sitter vacua we still have to live
with the cosmological constant problem.

\bigskip
\noindent {\large \bf Acknowledgements}\\[2ex]
I would like to thank my collaborators M.~Becker, G.~Curio and
moreover S.~Kachru, J.~Pati, E.~Poppitz, Q.~Shafi, G.~Shiu and
L.~Susskind for interesting discussions. Financial support came by
the National Science Foundation under Grant PHY-0099544.

 \newcommand{\zpc}[3]{{\sl Z. Phys.} {\bf C\,#1} (#2) #3}
 \newcommand{\npb}[3]{{\sl Nucl. Phys.} {\bf B\,#1} (#2) #3}
 \newcommand{\plb}[3]{{\sl Phys. Lett.} {\bf B\,#1} (#2) #3}
 \newcommand{\prd}[3]{{\sl Phys. Rev.} {\bf D\,#1} (#2) #3}
 \newcommand{\prb}[3]{{\sl Phys. Rev.} {\bf B\,#1} (#2) #3}
 \newcommand{\pr}[3]{{\sl Phys. Rev.} {\bf #1} (#2) #3}
 \newcommand{\prl}[3]{{\sl Phys. Rev. Lett.} {\bf #1} (#2) #3}
 \newcommand{\jhep}[3]{{\sl JHEP} {\bf #1} (#2) #3}
 \newcommand{\cqg}[3]{{\sl Class. Quant. Grav.} {\bf #1} (#2) #3}
 \newcommand{\prep}[3]{{\sl Phys. Rep.} {\bf #1} (#2) #3}
 \newcommand{\fp}[3]{{\sl Fortschr. Phys.} {\bf #1} (#2) #3}
 \newcommand{\nc}[3]{{\sl Nuovo Cimento} {\bf #1} (#2) #3}
 \newcommand{\nca}[3]{{\sl Nuovo Cimento} {\bf A\,#1} (#2) #3}
 \newcommand{\lnc}[3]{{\sl Lett. Nuovo Cimento} {\bf #1} (#2) #3}
 \newcommand{\ijmpa}[3]{{\sl Int. J. Mod. Phys.} {\bf A\,#1} (#2) #3}
 \newcommand{\rmp}[3]{{\sl Rev. Mod. Phys.} {\bf #1} (#2) #3}
 \newcommand{\ptp}[3]{{\sl Prog. Theor. Phys.} {\bf #1} (#2) #3}
 \newcommand{\sjnp}[3]{{\sl Sov. J. Nucl. Phys.} {\bf #1} (#2) #3}
 \newcommand{\sjpn}[3]{{\sl Sov. J. Particles \& Nuclei} {\bf #1} (#2) #3}
 \newcommand{\splir}[3]{{\sl Sov. Phys. Leb. Inst. Rep.} {\bf #1} (#2) #3}
 \newcommand{\tmf}[3]{{\sl Teor. Mat. Fiz.} {\bf #1} (#2) #3}
 \newcommand{\jcp}[3]{{\sl J. Comp. Phys.} {\bf #1} (#2) #3}
 \newcommand{\cpc}[3]{{\sl Comp. Phys. Commun.} {\bf #1} (#2) #3}
 \newcommand{\mpla}[3]{{\sl Mod. Phys. Lett.} {\bf A\,#1} (#2) #3}
 \newcommand{\cmp}[3]{{\sl Comm. Math. Phys.} {\bf #1} (#2) #3}
 \newcommand{\jmp}[3]{{\sl J. Math. Phys.} {\bf #1} (#2) #3}
 \newcommand{\pa}[3]{{\sl Physica} {\bf A\,#1} (#2) #3}
 \newcommand{\nim}[3]{{\sl Nucl. Instr. Meth.} {\bf #1} (#2) #3}
 \newcommand{\el}[3]{{\sl Europhysics Letters} {\bf #1} (#2) #3}
 \newcommand{\ap}[3]{{\sl Ann.~Phys.} {\bf #1} (#2) #3}
 \newcommand{\jetp}[3]{{\sl JETP} {\bf #1} (#2) #3}
 \newcommand{\jetpl}[3]{{\sl JETP Lett.} {\bf #1} (#2) #3}
 \newcommand{\acpp}[3]{{\sl Acta Physica Polonica} {\bf #1} (#2) #3}
 \newcommand{\sci}[3]{{\sl Science} {\bf #1} (#2) #3}
 \newcommand{\vj}[4]{{\sl #1~}{\bf #2} (#3) #4}
 \newcommand{\ej}[3]{{\bf #1} (#2) #3}
 \newcommand{\vjs}[2]{{\sl #1~}{\bf #2}}
 \newcommand{\hepph}[1]{{\sl hep--ph/}{#1}}
 \newcommand{\desy}[1]{{\sl DESY-Report~}{#1}}

\bibliographystyle{plain}

\end{document}